# Reduction of Phonon Lifetimes and Thermal Conductivity of a Carbon Nanotube on Amorphous Silica

Zhun-Yong Ong[1,2], Eric Pop[2,3,4,*] and Junichiro Shiomi[5,*]

[1]*Dept. of Physics, Univ. Illinois at Urbana-Champaign, Urbana IL 61801, USA*
[2]*Micro and Nanotechnology Lab, Univ. Illinois at Urbana-Champaign, Urbana IL 61801, USA*
[3]*Dept. of Electrical & Computer Eng., Univ. Illinois at Urbana-Champaign, Urbana IL 61801, USA*
[4]*Beckman Institute, Univ. Illinois at Urbana-Champaign, Urbana IL 61801, USA*
[5]*Dept. of Mechanical Eng., University of Tokyo, 7-3-1 Hongo, Bunkyo-ku, Tokyo. 113-8656, Japan*

We use molecular dynamics simulations to examine the phonon lifetimes in (10,10) carbon nanotubes (CNTs), both when isolated and when supported on amorphous $SiO_2$ substrates. We determine the Umklapp, normal, boundary and CNT-substrate phonon scattering rates from the computed inverse lifetimes. Suspended CNTs have in-plane optical phonon lifetimes between 0.7–2 ps, consistent with recent experiments, but contact with the substrate leads to a lifetime reduction to the 0.6-1.3 ps range. The thermal conductivity of the supported CNT is also computed to be ~30 percent lower than that of the isolated CNT. The thermal boundary conductance estimated from the CNT-substrate phonon scattering rates is in good agreement with that computed from the Green-Kubo relation and with previous experimental results. The results highlight that solid substrates can strongly affect and could be even used to tune the thermal properties of CNTs.

PACS numbers: 61.46.Fg, 65.80.-g, 63.20.kg, 68.35.-p

*Contact: epop@illinois.edu, shiomi@photon.t.u-tokyo.ac.jp

## 1. Introduction

Carbon nanotubes (CNTs) are a promising class of materials for nanoelectronic applications given their high mobility as well as their ability to sustain large current densities. However, at large current densities Joule heating becomes an important issue and the dissipation of waste heat must be taken into consideration to ensure optimal device performance [1-4]. Heat dissipation from the CNT depends on its thermal conductivity as well as the thermal boundary conductance of its interface with the environment [5, 6]. Furthermore, the microscopic nature of the heat generation adds to the complication of the dissipation problem. As hot electrons in the CNT are scattered, high energy optical phonons are emitted at an elevated rate, creating a non-equilibrium population of optical phonons, which in turn leads to an energy relaxation bottleneck and heat buildup within the CNT [7-9]. This highly non-equilibrium population of optical phonons also enhances electron scattering and lowers the current through the CNT [10-12]. Therefore, not only is it important to characterize thermal transport within and from the CNT to its environment but it is also especially relevant to develop a deeper understanding of the microscopic dynamics of phonon relaxation given their importance in high field transport. In this context, we use molecular dynamics (MD) simulations to study phonon lifetime changes induced by interaction with the substrate.

Although there have been numerous MD simulation-based studies of thermal conduction in isolated CNTs [13-17], relatively few works [18-23] study the effect of the environment on thermal conduction in CNTs. Even fewer are those in which the surrounding media is treated atomistically [18, 19, 23], partly because of the additional computational demands of doing so and partly because of the increased complexity of including external degrees of freedom. However, it is not clear that the neglect of the latter in the simulation of CNT thermal conductivity can be justified. When the CNT forms an interface with some surrounding medium or substrate, its phonons can be scattered by the static interfacial bonds (i.e. boundary scattering) as well as by dynamic coupling with the external vibrational modes (i.e. CNT-substrate phonon-phonon scattering). It is presently unclear which type of scattering mechanism is dominant, and therefore this is one of the questions we attempt to answer here.

In this paper, we use classical MD simulation and spectral energy density (SED) analysis [23-27] (a method formulated in Ref. [23], and further developed in this paper) to characterize the phonon lifetimes in an isolated (10,10) CNT and one that is supported by an amorphous silica (a-$SiO_2$) substrate. From the changes in phonon lifetimes as a result of contact with the substrate, the polarization, frequency and wavelength dependences of the substrate-induced perturbation are determined. We use the extracted phonon lifetimes to compute the thermal conductivity values of the isolated and supported CNTs. The reduction in the CNT thermal conductivity as a result of substrate contact is then related to the changes in the phonon lifetimes. By adjusting the simulation parameters, we are able to distinguish the different scatter-ing mechanisms arising from contact with the substrate and to estimate the boundary scattering and CNT-substrate phonon-phonon scattering rates for each CNT phonon mode.

The paper is organized as follows. The computational and data analysis methodologies are described in section 2. In particular, we show how the SED plot is computed from our MD simulation data and how the linewidths are extracted from the SED. In section 3, we present the linewidth data and analyze how the phonon lifetimes are affected by contact with the substrate. The scattering rates are resolved by type as CNT phonon-phonon (Umklapp/normal) scattering, boundary scattering and CNT-substrate phonon-phonon scattering. The relative contributions of the three processes to the phonon lifetimes are also quantified. The dependence of changes in phonon lifetimes are related to polarization, wavelength and frequency. In particular, we compute the phonon lifetime of the *G*-band phonons and determine the changes caused by contact with the substrate. In section 4, we use the scattering rates to compute the thermal conductivity of the isolated CNT and of the supported CNT. The relative contributions of the different acoustic and optical phonon branches are quantified in simulations with and without the substrate. We also calculate the thermal boundary conductance from the CNT-substrate phonon-phonon scattering rate and compare it with the value obtained from a Green-Kubo calculation.

# 2. Computational and Data Analysis Methodology

## A. MD Simulation Setup

Our MD simulations are performed using the LAMMPS package [28]. The structure consists of a 2000-atom, 50-unit cell long (10,10) CNT supported on an a-$SiO_2$ block with periodic boundary conditions in the plane of the interface. The setup of the simulation domain and structure essentially follows that in Refs. [29, 30] and is shown in Fig. 1. The CNT is oriented to be parallel to the *z*-axis and the surface of the substrate is parallel to the *x-z* plane. Periodic boundary conditions are imposed in the *x-z* directions. To model the C-C atomic interaction, we use the adaptive intermolecular reactive bond order potential [31]; to model the Si-Si, Si-O and O-O atomic interactions, we use the Munetoh [32] parameterization of the Tersoff potential [33]. The interaction between the CNT and the substrate atoms is assumed to be van der Waals (vdW). We model this vdW interaction with the Lennard-Jones (LJ) 12-6 potential $V_{ij}(r) = 4\varepsilon_{ij}[(\sigma/r)^{12}-(\sigma/r)^{6}]$ where the $\varepsilon_{ij}$ and $\sigma_{ij}$ parameters for $i$ = C and $j$ = Si, O are given in Ref. [29].

The temperature in all equilibrium MD simulations is 300 K. To reach the target temperature, the system is equilibrated at 300 K by first running it as an NVT ensemble for 20 ps before running it again as an NVE ensemble for 100 ps. We use a time step of 0.2 fs. During the NVT stage, we apply a Langevin thermostat to ensure that the phase space trajectory is randomized and that there are no ‘memory’ effects [34].

To elucidate the different scattering processes, we have three separate sets of simulations (cases I to III) with different conditions. As we progress from case I to III, more scattering mechanisms are included in the simulations. In the first set of simulations (case I), an isolated CNT is simulated at 300 K without any substrate. Thus, the phonon lifetimes are determined by only one scattering mechanism - the higher order (anharmonic) coupling between CNT phonons. In the second set of simulations (case II), the CNT is attached to the $SiO_2$ substrate. However, the substrate atoms are 'frozen' in their equilibrium positions and only the CNT atoms are allowed to move and interact with the frozen substrate atoms. By freezing the atoms in the substrate, the interaction between the CNT phonons and the substrate modes is prevented. Hence, in addition to the anharmonic coupling between CNT phonons, which we assume to be unchanged by the substrate, the CNT phonons are also affected by boundary scattering (i.e. perturbations from the static interfacial bonds with the substrate) but not by CNT-substrate phonon-phonon scattering. In the third set of simulations (case III), the substrate atoms are allowed to move so that interaction between the CNT phonons and the substrate modes can occur, resulting in CNT-substrate phonon-phonon scattering processes.

In all three sets of simulations, the individual atomic velocities of the 2000 CNT atoms are recorded at intervals of 8 fs for a total 655400 steps. The atomic velocities are then used to construct the spectral energy density (SED) as described in the following sub-section. A total of 15 independent runs [35] are performed for each set of simulations so that ensemble averages can be taken to reduce the noise in the SED analysis of the data.

## B. Spectral Energy Density Analysis

Although phonon lifetimes can be estimated from the time auto-correlation of the normal mode coordinates [19, 24, 36-38], such a method would be inconvenient in cases when the unit cell contains a large number of basis atoms. For example, in a (10,10) CNT, the 1D unit cell has 40 basis atoms, making it impractical to use the modal decomposition method. However, the SED method affords a computationally convenient approach to extract the phonon lifetimes in the CNT without having to decompose the atomic coordinates into their normal mode components. In the modal decomposition method, the normal modes have to be individually resolved and the lifetimes are then extracted from the individual time autocorrelations. On the other hand, the SED method converts the velocity data of groups of atoms into sets of data in frequency space and uses peak position and width analysis to determine the frequencies and lifetimes of the corresponding group of phonon modes with the same quantum numbers. In a sense, we can think of SED analysis as a kind of 'numerical spectroscopy'.

In order to extract the phonon lifetimes from the velocity history of the CNT atoms, we need to convert the spatial and time-dependent atomic velocity data into frequency and reciprocal space. This can be accomplished by taking advantage of the symmetries of the CNT. The (10,10) CNT has both translational symmetry in the axial direction and a 10-fold rotational symmetry in the circumferential direction. Given that the CNT is a one-dimensional crystal, it has a unit cell of 40 basis atoms with 120 phonon branches. We can label each of these unit cells with an index $n$ = 0 to 49 in our simulations as the CNT is 50 unit cells-long. However, because the symmetry of the unit cell is described by the $D_{10}$ point group, the translational unit cell can be decomposed into smaller translational-rotational unit cells with 4 basis atoms. Thus, we can label each of these smaller rotational unit cells within the translational unit cell with an index $n_\theta$. Each of the translational-rotational unit cells can be uniquely enumerated by the pair of indices ($n$,$n_\theta$) and thus each of the CNT atoms can thus be enumerated by a triplet of indices ($n$,$n_\theta$,$b$) where $b$ is the basis atom number. Therefore, as we move into the corresponding reciprocal spaces, the phonon modes can be enumerated by three quantum numbers: $k$ (the translational wave number from 0-49), $k_\theta$ (the rotational wave number from 0-9 with 0 corresponding to the circumferentially nodeless modes) and $\beta$ (the polarization index from 1-12). In this work, we enumerate the phonon modes by the indices ($k$,$k_\theta$,$\beta$).

We now describe the procedure for computing the SED from the atomic velocity data. The time derivative of the $\alpha$-th component ($r$, $\theta$ or $z$) of the $b$-th atomic coordinate in the unit cell $u_{b,\alpha}$ can be written as the sum of its modal components:

$$\dot{u}_{b,\alpha}(n,n_\theta,t)=\frac{1}{\sqrt{mNN_\theta}}\sum_{k=0}^{N-1}\sum_{k_\theta=0}^{N_\theta-1}\sum_{\beta=1}^{12}e_{b,\alpha}(k,k_\theta,\beta)\exp\left[-2\pi i\left(\frac{kn}{N}+\frac{k_\theta n_\theta}{N_\theta}\right)\right]\dot{q}(k,k_\theta,\beta,t) \tag{1}$$

where $m$ is the mass of the C atom, $N$ (= 50) is the number of translational unit cells, $N_\theta$ (= 10) is the number of rotational unit cells, $e_{b,\alpha}$ is the eigenvector component of the ($k$,$k_\theta$,$\beta$) mode and $q$ is the normal coordinate of the latter. Following equation (4) of Ref. [23], the phonon spectral energy density is

$$\Phi(k,k_\theta,\nu)=\left\langle\sum_{b=1}^{4}\sum_{\alpha}\frac{m}{2}\left|\frac{1}{NN_\theta\tau}\sum_{n=0}^{N-1}\sum_{n_\theta=0}^{N_\theta-1}\exp\left[2\pi i\left(\frac{kn}{N}+\frac{k_\theta n_\theta}{N_\theta}\right)\right]\int_0^\tau dt\,\exp(2\pi i\nu t)\dot{u}_{b,\alpha}(n,n_\theta,t)\right|^2\right\rangle \tag{2}$$

which simplifies to

$$\Phi(k,k_\theta,\nu)=\frac{1}{2}\left\langle\left|\frac{1}{\tau}\int_0^\tau dt\,\exp(2\pi i\nu t)\sum_{\beta=1}^{12}\dot{q}(k,k_\theta,\beta,t)\right|^2\right\rangle \tag{3}$$

where $\nu$ is the frequency. Our expression for the phonon SED in Eq. (2) only differs from equation (4) of Ref. [23] by a prefactor of 1/(4π) and its additional use of the rotational wave vector. By summing Eq. (3) with respect to $k_\theta$, we can recover equation (4) in Ref. [23].

Using the Wiener-Khinchine theorem, which implies that the absolute square of the Fourier transform is equal to the Fourier transform of the autocorrelation [34], the expression in Eq. (3) simplifies to

$$\Phi(k,k_\theta,\nu)=\frac{1}{2\tau}\int_0^\tau dt\,\exp(2\pi i\nu t)\sum_{\beta=1}^{12}\left\langle \dot{q}(k,k_\theta,\beta,t)^*\,\dot{q}(k,k_\theta,\beta,0)\right\rangle \tag{4}$$

i.e. the Fourier transform of the sum of the normal mode time autocorrelations. If we assume that the $(k,k_\theta,\beta)$ normal mode has an inverse lifetime of $\gamma(k,k_\theta,\beta)=1/\tau(k,k_\theta,\beta)$ and an eigenfrequency of $\nu(k,k_\theta,\beta)$, then we can write the time autocorrelation function as an exponentially decaying cosine function

$$\left\langle \dot{q}(k,k_\theta,\beta,t)^*\,\dot{q}(k,k_\theta,\beta,0)\right\rangle \approx \left\langle \left|\dot{q}(k,k_\theta,\beta,0)\right|^2\right\rangle \cos\left(2\pi\nu(k,k_\theta,\beta)t\right)\exp\left(-\gamma(k,k_\theta,\beta)t/2\right) \tag{5}$$

Hence, after inserting Eq. (5) back into Eq. (3) and restricting ourselves to only positive frequencies, we obtain the expression for $\Phi(k,k_\theta,\nu)$ as a sum (over all polarizations) of Lorentzian functions i.e.

$$\begin{aligned}\Phi(k,k_\theta,\nu)&=\frac{2}{\tau}\sum_{\beta=1}^{12}\frac{\gamma(k,k_\theta,\beta)\left\langle\left|\dot{q}(k,k_\theta,\beta,0)\right|^2\right\rangle}{16\pi^2\left[\nu-\nu(k,k_\theta,\beta)\right]^2+\gamma(k,k_\theta,\beta)^2}\\&=\sum_{\beta=1}^{12}\frac{I(k,k_\theta,\beta)}{16\pi^2\left[\nu-\nu(k,k_\theta,\beta)\right]^2+\gamma(k,k_\theta,\beta)^2}\end{aligned} \tag{6}$$

where $I(k,k_\theta,\beta)$ is the intensity of the phonon mode $(k,k_\theta,\beta)$.

Here, for the $(k,k_\theta,\beta)$ phonon mode, we point out that the inverse phonon lifetime $\gamma$ in Eq. (5) is the average of the phonon *population* inverse lifetime $1/\tau_{\text{pop}}$ and the pure *dephasing* inverse lifetime $1/\tau_{\text{dep}}$ [39, 40], i.e. $2\gamma = 1/\tau_{\text{pop}} + 1/\tau_{\text{dep}}$. The former ($\tau_{\text{pop}}$), also known as the relaxation time, is the actual quantity of interest especially when it comes to the computation of thermal transport coefficients, although most papers that attempt to compute such coefficients from the mode lifetimes do not appear to draw this distinction (for example see Refs. [23, 36, 37]). For general theoretical completeness, we state explicitly our assumption that $\tau_{\text{pop}} = \tau_{\text{dep}}$ and thus $\tau_{\text{pop}} = 1/\gamma$. This assumption can be at least justified in the oscillator-bath framework [39, 40] by considering the phonon mode to be the harmonic oscillator and the other phonon modes (CNT and substrate) to form the thermal bath to which it is weakly coupled. It is also necessary if we are to compare our computed CNT phonon lifetimes to the experimentally measured phonon population lifetimes in the literature (for example see Refs. [8, 41]). In the rest of the text, we will use the terms phonon lifetime and relaxation time interchangeably.

In Fig. 2(a), we plot $\Phi(k,k_\theta,\nu)$ summed over all $k$ and $k_\theta$ values. The plot is proportional to the phonon density of states at 300 K and comprises 6000 peaks clustered closely together, making it unfeasible to fit the width of the individual peaks. In Fig. 2(b), we plot $\Phi(k,k_\theta,\nu)$ summed over all $k_\theta$ values for $k = 0$. We have a total of 120 peaks and it is now possible to observe some of the individual peaks. However, especially in the $\nu$ = 47-55 THz region, many of the peaks are still closely packed together. In Fig. 2(c), $\Phi(k,k_\theta,\nu)$ corresponding to $k = 0$ and $k_\theta = 0$ is plotted. This time, we have 12 distinct and well-separated peaks and the problem of closely clustered peaks in the $\nu$ = 47-55 THz region is solved. The peaks are sufficiently separated that we can identify the split G (G- and G+) peaks in Fig. 2(d).

For each wave vector in the $(k,k_\theta)$-space, the frequency-dependent SED spectrum has 12 peaks [see Eq. (6)] and was fitted by multiple Lorentzian functions using the Levenberg-Marquardt algorithm [42]. The fitting algorithm requires an initial guess of the Lorentzian parameters, which was determined by using a peak detection algorithm [42] knowing the number of peaks in a spectrum. Since the peak detection algorithm suffers from noise in the spectra, the peak assignments were manually checked. The process could be fully automated by using the phonon frequencies calculated from harmonic lattice dynamics as the initial guess, however this was not performed in the current work since the influence of the substrate on the CNT phonon frequencies was not known in advance. Nevertheless, by the quasi-automatic procedure, SED spectra could be fitted with sufficiently small nominal residual (the weighted sum of squared residuals was less than 10 percent of that of the signal).

For the CNT supported on a-$SiO_2$ substrate, the CNT phonon frequencies are expected to shift from those of the isolated CNT due to the interaction with the substrate. However, for the current system with weak vdW interaction between the CNT-substrate, equilibrium MD simulations find the changes induced by the substrate in the CNT phonon dispersion to be minute for most of the modes. In other words, the perturbation caused by the substrate does not appreciably influence the harmonic properties of the CNT. Therefore, for the sake of simplicity in the phonon lifetime analysis to follow, we have ignored the change in eigenvalues and eigenstates and assigned the relaxation time $\tau(k,k_\theta,\beta)$ of the supported CNT to the original unperturbed eigenvalues and eigenstates of the isolated CNT with nearest distance in the $(\nu,k,k_\theta)$ space. During this analysis, we have ignored the inter-CNT-substrate modes, since the non-propagating modes with limited degrees of freedom have negligible contribution to the thermal transport properties, even though their frequencies are smaller and relaxation times are longer than the lowest frequency CNT phonon.

## 3. Phonon Lifetime Analysis

We assume that only three mechanisms contribute to the CNT phonon lifetime in our MD simulations:

1. Anharmonic coupling or Umklapp/Normal scattering between CNT phonon modes ($1/\tau_U$)
2. Boundary scattering by static vdW bonds at the interface ($1/\tau_B$)
3. CNT-substrate phonon-phonon scattering ($1/\tau_S$)

If we assume that these processes are independent, we can invoke Matthiessen's rule and write the inverse lifetime of the $(k,k_\theta,\beta)$ phonon mode as a sum of inverse scattering times

$$\gamma(k,k_\theta,\beta)=\frac{1}{\tau(k,k_\theta,\beta)}=\frac{1}{\tau_U(k,k_\theta,\beta)}+\frac{1}{\tau_B(k,k_\theta,\beta)}+\frac{1}{\tau_S(k,k_\theta,\beta)} \tag{7}$$

As all three processes contribute to the phonon lifetime, we need to modify the simulation conditions to isolate their effects. To do that, three different sets of MD simulations are performed. In the first set (case I), the substrate is removed from the simulation setup, leaving only the CNT. For notational convenience, we suppress the phonon index $(k,k_\theta,\beta)$ in the following discussion and whenever it is obviously not needed. The inverse phonon lifetime $\gamma_I$ extracted from these simulations is given by $\gamma_I = 1/\tau_U$. In the second set (case II), the substrate is put back into the simulation setup but its atoms are frozen. This freezing of the substrate atoms means that there is no CNT-substrate phonon-phonon scattering. Hence, the inverse phonon lifetime $\gamma_{II}$ from these simulations gives us $\gamma_{II} = 1/\tau_U + 1/\tau_B$. In the third set (case III), the substrate atoms are no longer frozen and we run everything at 300 K. Therefore, we have $\gamma_{III} = 1/\tau_U + 1/\tau_B + 1/\tau_S$. Given the three inverse phonon lifetimes $\gamma_I$, $\gamma_{II}$ and $\gamma_{III}$, it is straightforward to work backwards to reconstruct the inverse scattering times: $1/\tau_U = \gamma_I$, $1/\tau_B = \gamma_{II} - \gamma_I$ and $1/\tau_S = \gamma_{III} - \gamma_{II}$.

It must be mentioned that underlying our procedure of modifying the simulation conditions to isolate the individual scattering rates is the assumption that the scattering rates $1/\tau_U$, $1/\tau_B$ and $1/\tau_S$ are not significantly altered by these modifications. For example, we have assumed that the $1/\tau_U$ term is unchanged in $\gamma_I$, $\gamma_{II}$ and $\gamma_{III}$. This assumption should hold provided that the CNT-substrate interaction is sufficiently weak. However, this 'sufficiently weak' assumption can only be justified *post hoc* from our simulation results and analyses in the following subsections.

## A. Phonon Lifetimes in Isolated CNT

Figure 3 shows the frequency dependence of phonon relaxation times of the isolated CNT which corresponds to experiments as in Refs. [10, 43, 44]. The general trend agrees with those reported for other materials such as silicon [45, 46]; stronger and monotonic frequency dependence in the low frequency regime, and weaker and non-monotonic dependence in the high frequency regime. The acoustic phonon branches show particularly strong frequency dependence, where the relaxation time varies by approximately two orders of magnitude for an order of magnitude variation in frequency. The phonon-branch dependence of the relaxation time is weaker among the acoustic phonons than between acoustic and optical phonons in the low frequency regime, suggesting stronger dependence on the circumferential wave number than on the polarization.

The trend of the frequency dependence is compared with the well known scaling $\tau \propto \nu^{-2}$ obtained by Klemens [47] for three-phonon scattering of linear-dispersion modes at low frequency and high temperature limits. The same scaling law for acoustic modes in CNTs was also obtained by Hepplestone and Srivastava [48], also using three-phonon scattering. Although statistical error prevents us from accurately estimating the effective exponent, the twisting acoustic (TW) modes with linear dispersion exhibit a power-law frequency dependence with an exponent (~ -1.1) clearly different from -2, thus having weaker frequency dependence. This could be due to higher order anharmonic events, which are not taken into account in Klemens' scaling, or due to the one-dimensional nature of the CNT although the result from Refs. [48] and [49] seems to exclude that possibility. The degenerate transverse acoustic (TA) modes do not show a single power law trend presumably due to the nonlinear dispersion caused by the CNT flexure modes [50]. The exponent of the longitudinal acoustic (LA) modes could not be identified due to the limited CNT length. Note that, for the LA mode, the finite length of the current system gives only two data points in the frequency range with linear dispersion (< 3 THz) [25].

## B. Changes in Phonon Lifetimes from Substrate Contact

We next consider phonon relaxation in CNT devices supported by a-$SiO_2$ substrates which correspond to experiments in Refs. [5, 6, 51]. In this case, the SED analysis finds that the reduction of CNT phonon lifetimes depends strongly on the phonon states. The reduction is most noticeable in the low frequency phonons. Strong polarization dependence is also observed, where the reduction of the relaxation time is the largest for transverse modes. Note that the optical phonons with frequency smaller than 4 THz mostly have transverse polarization. The actual scattering rate due to the substrate and its frequency dependence can be better quantified by calculating the substrate scattering rate $\gamma_{SiO2} = \gamma_{III} - \gamma_{I}$. Figure 4 shows that the $\gamma_{SiO2}$ spectrum has a broad peak in the low frequency regime up to 10 THz, and sharper peaks in the regimes around 18 THz and 50 THz. In contrast, $\gamma_{SiO2}$ is nearly zero in the intermediate frequency regime between 18 THz and 50 THz. Such selectivity comes from the strong branch dependence of the substrate scattering as clearly demonstrated by the contour of $\gamma_{SiO2}$ in the $(\nu,k)$-space [Fig. 5(a)]. Furthermore, this was found to correlate well with the magnitude of the radial atomic displacements of the CNT eigenvectors [Fig. 5(b)], and therefore, the selectivity originates from a rather intuitive mechanism that the substrate effectively scatters CNT phonons with radial atomic displacements.

Based on the $\gamma_{II}$ obtained from the simulation with the *frozen* substrate, we can separate the scattering due to the substrate into boundary scattering ($1/\tau_B$) and phonon-phonon scattering ($1/\tau_S$), i.e. $\gamma_{SiO2} = 1/\tau_B + 1/\tau_S$. As seen in Fig. 4, where the difference between $\gamma_{III}$-$\gamma_{I}$ ($=1/\tau_B+1/\tau_S$) and $\gamma_{II}$-$\gamma_{I}$ ($=1/\tau_B$) is small in the entire frequency domain, the reduction in CNT phonon lifetime is dominantly caused by boundary scattering. This means that the dynamics of the a-$SiO_2$ substrate has negligible influence on the intrinsic pho-non relaxation of CNT. This is consistent with the above observation that the mode selectivity of the substrate scattering rate is determined by the eigenvectors of the CNT but not by those of the $SiO_2$ substrate. Therefore, the influence of the substrate on the intrinsic heat conduction of CNT can be sufficiently described by a CNT with static perturbation of the potential, which simplifies the physical model to investigate reduction of heat conduction or enhancement of hot phonon dissipation.

### C. G-band Phonon Lifetime

The CNT optical phonons at the Γ point (*G*-mode phonons), which correspond to the C-C bond stretching motion, are known to couple strongly with electrons [6, 10, 12, 52]. In high-field transport, these high-frequency phonons play an important role in the energy relaxation of hot electrons. Hence, the decay dynamics and lifetimes of these phonons are expected to play an important role in electrical transport.

The lifetimes of the in-plane optical phonons in the $v = 47$ to 55 THz range, where the *G*-mode phonons are in our simulations [53], for cases I (no substrate) and III (with substrate) are shown in Fig. 6. The frequency range between 49.5-54.2 THz contains the *G*-mode phonons as well as other high-frequency longitudinal optical (LO) and in-plane transverse optical (TO) modes between the Γ and *K* points. In general, the phonons in this range have lifetimes between $\tau = 0.7$–$2.0$ ps for case I, consistent with the experimentally measured *G*-mode phonon population lifetimes (~1.1±0.2 ps) in Refs. [8, 41].

To make the direct connection to experimental measurements [41], we identify the Raman-active *G* phonons, making use of the point group symmetry of the unit cell. Since the $k_\theta = 0$ spectrum conforms to the circumferentially nodeless state, its peaks have $A_1$ symmetry. In the $k = 0$ and $k_\theta = 0$ spectrum [Fig. 2(c)], there are 12 distinct peaks, two of which are the *G*- and *G*+ phonon modes [see Fig. 2(d)]. The identification of the *G*- and *G*+ phonon modes is made on the basis of polarization and symmetry consideration. When the spectrum in Fig. 2(d) is recomputed with only transversely polarized components, we find that the high-frequency peaks ($v = 49.6$ and 54.1 THz) on either side of the $v = 52.6$ THz peak disappear, indicating that they are longitudinally polarized and that the $v = 52.6$ THz peak is transversely polarized. The $A_1$-symmetry *G* peaks correspond to the Γ-point in-plane TO (iTO) and the LO phonon modes, and are formed from the splitting of the *G* peak in graphene as a result of curvature effects which lead to the softening of the iTO mode with respect to the LO mode. Thus, we identify the $v = 52.6$ THz mode as the *G*- (iTO) peak and the $v = 54.1$ THz mode as the *G*+ (LO) peak. It should be pointed out that, when electron-phonon coupling effects are taken into account in a metallic CNT, the order of identification is reversed, i.e. the *G*- peak corresponds to the LO mode and vice versa as a result of LO phonon softening from the Kohn anomaly effect [54]. However, in a classical MD simulation, these effects are absent and we attribute the splitting purely to curvature effects.

We fit the peaks and find that the *G*- and *G*+ lifetimes are $1/\gamma_{I}$ = 1.00 and 0.99 ps respectively in the isolated CNT and $1/\gamma_{III}$ = 0.62 and 0.69 ps in the supported CNT. Their respective scattering rates are $\gamma_{SiO2}$ = 0.61 and 0.44 THz. Unsurprisingly, being transversely polarized, the *G*- (iTO) mode is more strongly scattered than the *G*+ (LO) mode by the substrate. Nonetheless, the lifetime of the *G*+ (LO) mode, which is coupled strongly to electrons in real systems, is still reduced significantly by ~30 percent. Even when the substrate is 'frozen' the *G*+ peak lifetime reduction is almost the same, indicating that it is due mainly to boundary scattering and that the *G*+ phonons do not couple directly with the modes in the substrate.

In high field electrical transport in CNTs, Γ-point LO phonons are preferentially emitted by hot electrons, leading to an elevated non-equilibrium LO phonon population [4, 9, 10, 12, 52]. Intra-nanotube coupling with lower frequency phonons can provide channels for the LO phonons to decay and dissipate energy although for freely suspended CNTs [10] these intra-nanotube channels are insufficient and lead to an energy relaxation bottleneck. From our MD simulation results, we find that boundary scattering in CNTs on a-$SiO_2$ provides additional intra-nanotube decay channels for the high-frequency phonons. In the context of high-bias electrical transport, this phonon lifetime reduction may ameliorate the energy relaxation bottleneck [1, 10, 55].

In Ref. [41], the *G*+ lifetime was measured to be around 1.1 ± 0.2 ps for a semiconducting (6,5) CNT, in good agreement with our MD simulation results ($1/\gamma_{I}$ = 0.99 ps). However, no significant changes to the *G*+ mode phonon lifetimes were found when the CNT was immersed in liquid $D_2O$ using two different types of surfactants, somewhat contradicting the ~30 percent lifetime reduction from our MD simulations. However, our simulations involve a solid substrate while their experiments were performed with CNTs in a liquid suspension. The perturbation due to the a-$SiO_2$ substrate is likely different from that by the surfactants and hence this discrepancy is not entirely unexpected.

## 4. Thermal Transport Coefficients

### A. Thermal Conductivity of Isolated and Supported CNTs

By knowing the mode-dependent phonon relaxation time, the thermal conductivity in the classical limit can be calculated as

$$\kappa = \frac{1}{V}\sum_{k_\theta}\sum_{k}\sum_{\beta=1}^{12} k_B \upsilon_{\beta,k,k_\theta}^2 \tau_{\beta,k,k_\theta} \tag{8}$$

where $k_B$ is the Boltzmann constant. For the CNT volume, we adopt a conventional definition $V = \sqrt{3}\pi a_c N b_c d$, where $a_c$, $b_c$, and $d$ are the lattice constant, vdW distance (3.4 Å), and CNT diameter, respectively. The group velocity $\upsilon$ was extracted from the SED at 300 K, and hence, includes the anharmon-ic effects. The obtained thermal conductivity of the isolated CNT was $\kappa_{iso} \approx 475$ $Wm^{-1}K^{-1}$ and the four acoustic branches carry *only* 40% of the total heat current. These values are in reasonable agreement with those reported by Thomas *et al.* [23], where the counterintuitively small contribution of acoustic phonons was attributed to the appreciable contribution from the low frequency optical phonons. For the CNT on a-$SiO_2$ substrate, thermal conductivity is reduced by 33 percent to $\kappa_{susp} \approx 317$ $Wm^{-1}K^{-1}$. As summarized in Table 1, more than 70 percent of the thermal conductivity reduction is attributed to the acoustic phonons. The thermal conductivity reduction in each mode is 83%, 51%, and 35% for TA, LA, and TW phonons, which are significantly larger than the reduction in the rest of the modes (~16%). The largest reduction is for the TA modes, consistent with the correlation between the substrate scattering rate and the magnitude of transverse component eigenvectors discussed in Section 3B. As a consequence, in the supported CNT, the acoustic phonons carry an even smaller fraction (25%) of the total heat conducted than in the isolated CNT. This demonstrates the strongly multi-phonon-band nature of CNT heat conduction particularly when a CNT is supported by a substrate.

It is worth commenting on the size effect of thermal conductivity, i.e. the dependence of thermal conductivity on the supercell length when the length is smaller than phonon mean-free-path. Thomas *et al.* [23] performed simulations for different CNT lengths and observed convergence of thermal conductivity at the same length with the current study (50 unit cells). One justification is that a periodic supercell calculation, unlike the *real* finite length system [56], allows the phonons to travel a much longer distance than the computational cell through the periodic boundary, and thus models an infinite length system. On the other hand, this ignores the phonons with wavelength longer than the supercell length that might have contributed to the *bulk* thermal conductivity. Although in the current work we rely on the work of Thomas *et al.* [23], it is not evident at the moment why the size dependence should diminish at a supercell size much smaller than phonon mean-free-paths.

## B. Thermal Boundary Conductance from Substrate Phonon Scattering

In the previous section, we computed the CNT-substrate phonon-phonon scattering rates, $\tau_S(k,k_\theta,\beta)^{-1}$. If our assumption that these are the relaxation rates at which the CNT phonons scatter with the substrate phonons is correct, then it is possible to estimate the interfacial thermal transport coefficient [5, 29]. The expression for the interfacial heat flux between CNT phonons and the substrate modes is

$$Q = \sum_k \sum_{k_\theta} \sum_{\beta=1}^{12} h\nu(k,k_\theta,\beta)\tau_S(k,k_\theta,\beta)^{-1} \frac{dn(k,k_\theta,\beta)}{dT} \Delta T = gL\Delta T \quad (9)$$

where $\nu(k,k_\theta,\beta)$, $\tau_S(k,k_\theta,\beta)$ and $n(k,k_\theta,\beta)$ are the frequency, the inverse CNT-substrate phonon-phonon scattering rate and the Bose-Einstein occupation factor of the $(k,k_\theta,\beta)$ mode respectively; $\Delta T$ is the tempera-ture differential between the CNT and the substrate, $g$ is the thermal boundary conductance and $L$ (= 122.8 Å) is the length of the CNT. Thus, the CNT-substrate thermal boundary conductance (TBC) per unit length is

$$g = \frac{1}{L}\sum_{k}\sum_{k_\theta}\sum_{\beta=1}^{12} h\nu(k,k_\theta,\beta)\tau_S(k,k_\theta,\beta)^{-1}\frac{dn(k,k_\theta,\beta)}{dT} \tag{10}$$

and in the classical limit, where $h\nu(k,k_\theta,\beta)\times dn(k,k_\theta,\beta)/dT \approx k_B$, the expression for the TBC reduces to

$$g = \frac{k_B}{L}\sum_{k}\sum_{k_\theta}\sum_{\beta=1}^{12}\tau_S(k,k_\theta,\nu)^{-1} = \frac{Nk_B}{L}\left\langle\frac{1}{\tau(k,k_\theta,\nu)}\right\rangle \tag{11}$$

where $\langle\ldots\rangle$ is the average taken over all modes and $N$ is the total number of phonon modes. In that case, it is relatively straightforward to compute $g$ since

$$N\left\langle\frac{1}{\tau_S(k,k_\theta,\beta)}\right\rangle = \sum_{k}\sum_{k_\theta}\sum_{\beta=1}^{12}\left[\gamma_{III}(k,k_\theta,\beta)-\gamma_{II}(k,k_\theta,\beta)\right] \tag{12}$$

In other words, the TBC is simply proportional to the difference in the sum of the inverse phonon lifetimes between case II and III. From the formula in Eq. (11), we compute the TBC to be $g \approx 0.055$ $WK^{-1}m^{-1}$ for this (10,10) CNT on $SiO_2$, consistent with our previous findings [5, 29]. We also compute the TBC using the Green-Kubo relation given in Ref. [30]

$$g_{GK} = \frac{1}{Lk_BT^2}\int_0^\infty dt\,\langle J(t)J(0)\rangle \tag{13}$$

where $J$ is the atomistic interfacial heat flux between the CNT and the substrate and $\langle\ldots\rangle$ here refers to the ensemble average. The same equilibrium MD simulation setup as shown in Fig. 1 was run at 300 K for 1 ns. From this simulation, we compute the autocorrelation of $J$ and use it to determine $g_{GK}$. We estimate $g_{GK} = 0.075\pm0.018$ $WK^{-1}m^{-1}$ which is in reasonable agreement with the value of $g$ we computed from the CNT-substrate phonon-phonon scattering rates. The reasonable agreement between the two methods of computing the TBC suggests that our estimate of $1/\tau_S(k,k_\theta,\beta)$ is at least not wildly inaccurate and validates our earlier assumptions in Eq. (7) that the scattering mechanisms are independent and the CNT-substrate interaction is sufficiently weak. This result also confirms that CNT-substrate phonon-phonon scattering plays only a minor role in the reduction of the CNT phonon lifetimes because the TBC would be larger if the CNT-substrate phonon-phonon scattering were stronger.

To see the frequency dependence of the TBC, we plot in Fig. 7 the differential TBC $g'(\nu)$

$$g'(\nu) = \frac{k_B}{L} \sum_{k} \sum_{k_\theta} \sum_{\beta=1}^{12} \frac{1}{\sqrt{4\pi^2 \Delta\nu}} \exp\left[ -\frac{[\nu - \nu(k, k_\theta, \beta)]^2}{2(\Delta\nu)^2} \right] \tau_S(k, k_\theta, \beta)^{-1} \tag{14}$$

which convolves a Gaussian function and the scattering rate, with the broadening parameter $\Delta\nu$ = 0.01 THz. The TBC can be obtained by integrating $g'$ over the frequency range. We find that $g'$ is the largest in the 0–9 THz range, suggesting that the modes in this range dominate interfacial thermal transport. In certain frequency ranges, especially between 17.7–19.4 THz, $g'$ has negative values which come about when $\gamma_{\mathrm{II}}$ is greater than $\gamma_{\mathrm{III}}$. Another region where we can find large negative values is between 50.5–55 THz. The negative values can either mean that the CNT-substrate phonon-phonon scattering rate is negative, which is unphysical, or more probably that the true $1/\tau_S$ cannot be accurately determined from the difference in the $\gamma_{\mathrm{III}}$ and $\gamma_{\mathrm{II}}$. Nonetheless, when we integrate $g'(\nu)$ from 0 to 9 THz, we find a value that is ~98 percent of $g$ (= 0.055 $\mathrm{WK^{-1}m^{-1}}$). The overall picture strongly supports the idea that interfacial thermal transport is dominated by the low frequency CNT phonons, a conclusion also reached in Refs. [30, 57]. In particular, in Ref. [30], where the simulation set up was almost identical, it was found that in a heat-pulsed (10,10) supported on an a-$SiO_2$ substrate, the sub-10 THz, long wavelength phonons underwent more rapid temperature relaxation than the rest of the CNT.

## 5. Conclusions

From the SED analysis of our classical MD simulation of a (10,10) CNT on an a-$SiO_2$ substrate, our main findings are

a. the different scattering rates (anharmonic, boundary, CNT-substrate phonon-phonon) can be distinguished by changing the simulation conditions,
b. the $G$+ (Γ-LO) phonon mode undergoes a ~30 percent lifetime reduction between suspended and $SiO_2$-supported CNTs, suggesting that the substrate reduces the energy relaxation bottleneck,
c. the thermal conductivity is also reduced by ~30 percent between suspended and supported CNTs mainly due to boundary scattering,
d. CNT-substrate phonon-phonon scattering rates are much smaller than boundary scattering rates,
e. interfacial thermal transport is dominated by phonons in the 0–9 THz range, and
f. the value of the thermal boundary conductance can be extracted from the CNT-substrate phonon-phonon scattering rates and agrees with that computed from Green-Kubo calculations.

From the perspective of heat dissipation in CNT-based electronics, the interaction between the CNT and the substrate plays a complex role. While it leads to a degradation of the CNT thermal conductivity, it also provides an additional channel for energy dissipation via thermal boundary conduction into the sub-strate. Because the dominant mechanisms responsible for the thermal conductivity degradation and interfacial thermal transport are different, this suggests that the degradation in the CNT thermal conductivity could be mitigated without affecting interfacial thermal transport, through chemical or physical modification of the substrate surface. We surmise that a more pristine and smoother surface will reduce the number of external static scattering sites and lead to a smaller decrease in thermal conductivity without reducing the thermal boundary conductance.

## Acknowledgement

Part of this work (Z.-Y. Ong and E. Pop) was supported by the Nanoelectronics Research Initiative (NRI) SWAN center, the NSF under Grant No. CCF 08-29907, and a gift from Northrop Grumman Aerospace Systems (NGAS). Part of this work (J. Shiomi) was supported by KAKENHI 23760178.

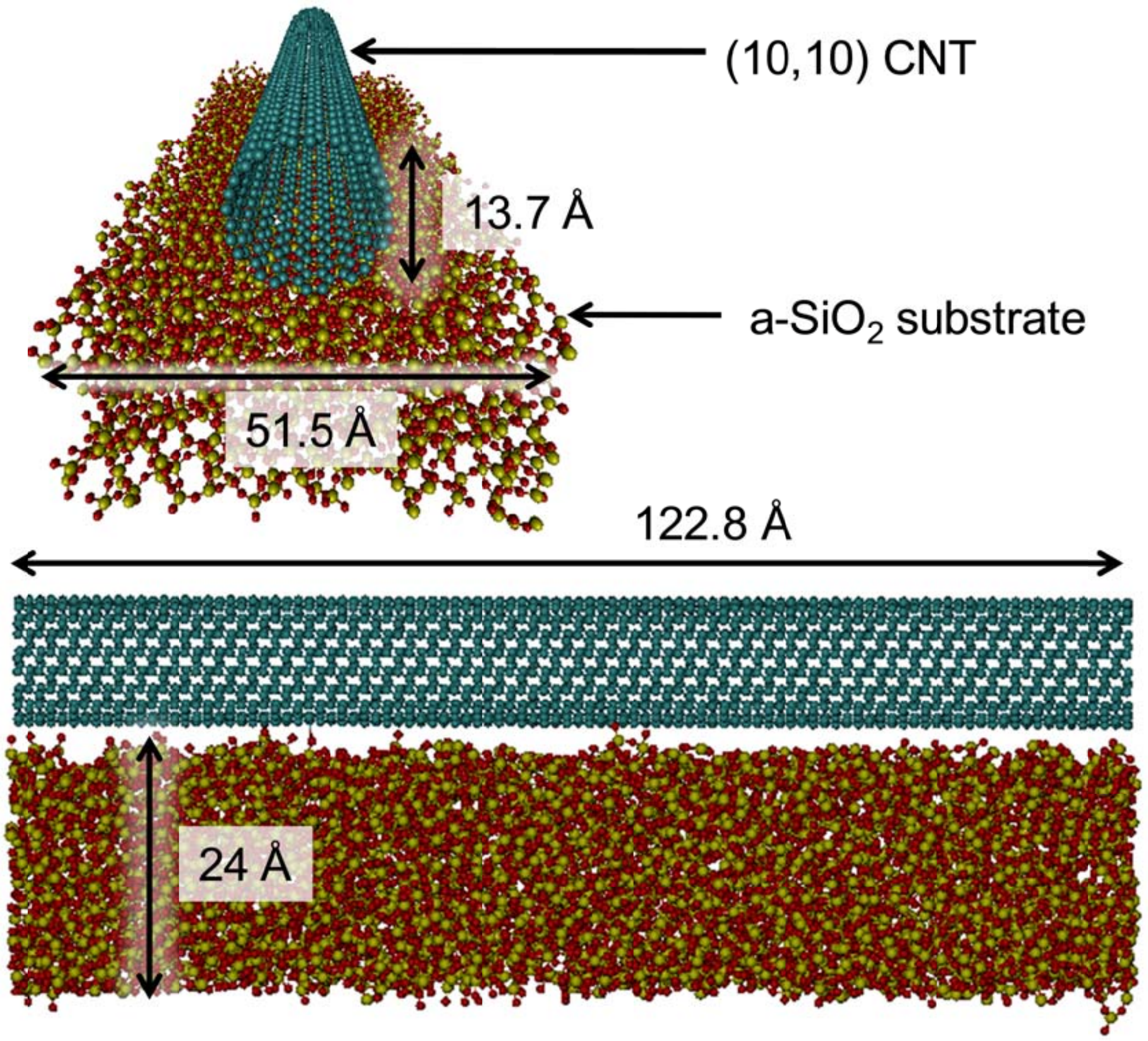

**Figure 1**: (Color online) Rendering of the simulation setup with a (10,10) CNT supported on an a-$SiO_2$ block mimicking the substrate. The a-$SiO_2$ block is 24 Å thick, 51.5 Å wide and 122.8 Å long. The CNT is 122.8 Å (50 unit cells) long and has a diameter of about 13.7 Å. **(Top)** Cross-sectional view of the set-up. **(Bottom)** Side view of the setup. The surface of the a-$SiO_2$ substrate is relatively flat with some surface roughness. Periodic boundary conditions are imposed in the plane parallel to the substrate surface.

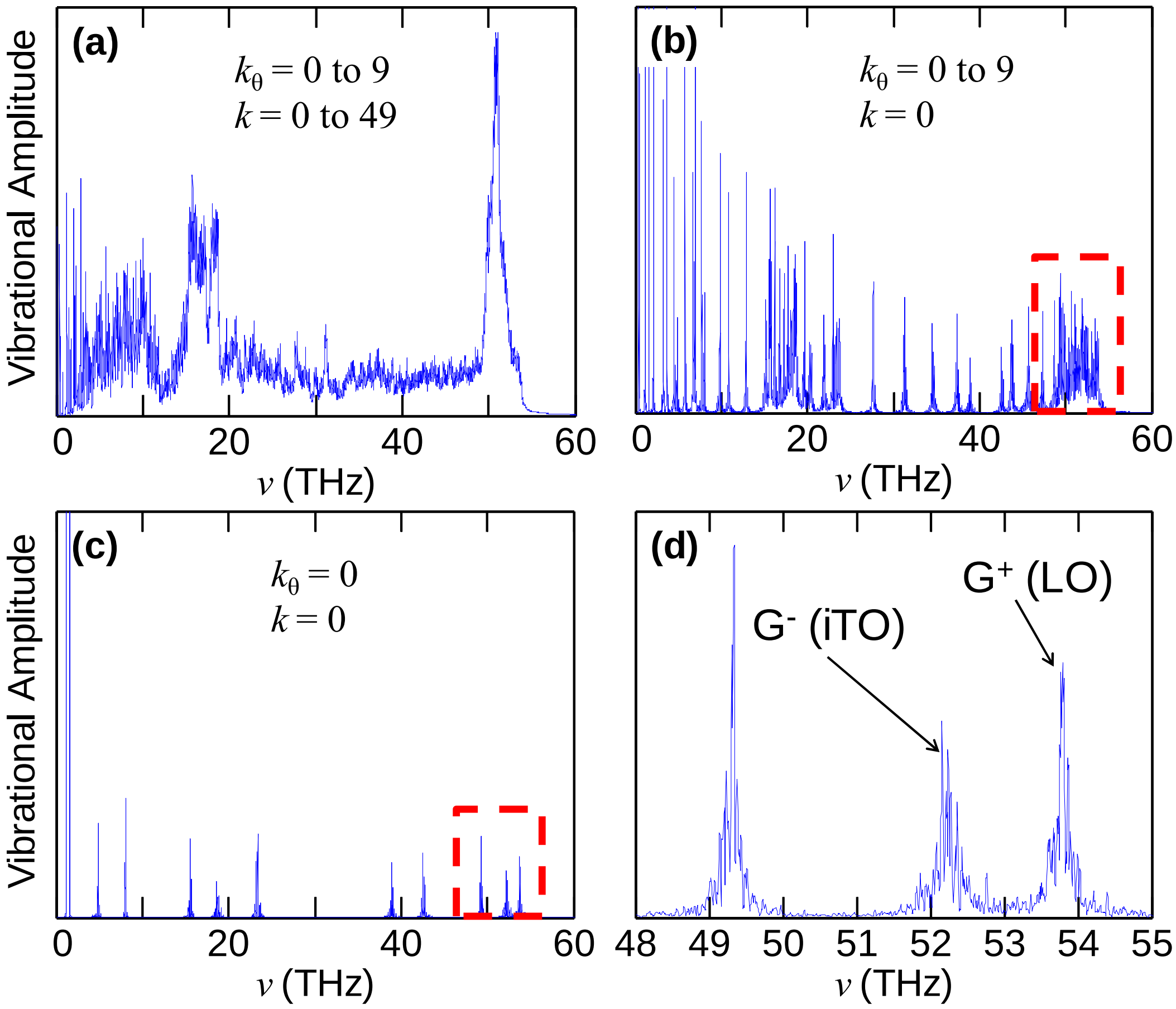

**Figure 2:** (Color online) We plot $\Phi(k,k_\theta,v)$ over all $k$ and $k_\theta$ values for $T$ = 300 K in **(a)**. Given that there are 6000 peaks, we cannot distinguish between the individual peaks. However, by restricting ourselves to a particular $k$, we reduce the number of peaks to 120, as shown in **(b)** where $k$ = 0. Even so, many of the peaks are still closely clustered together, especially in the 47 to 55 THz region (enclosed by red dashed lines). In **(c)**, we restrict $k_\theta$ = 0, leading to only 12 distinct peaks, and resolving the problem of the closely clustered peaks between 47 and 55 THz. **(d)** The three peaks are sufficiently separated, enabling us to distinguish the G- and G+ modes.

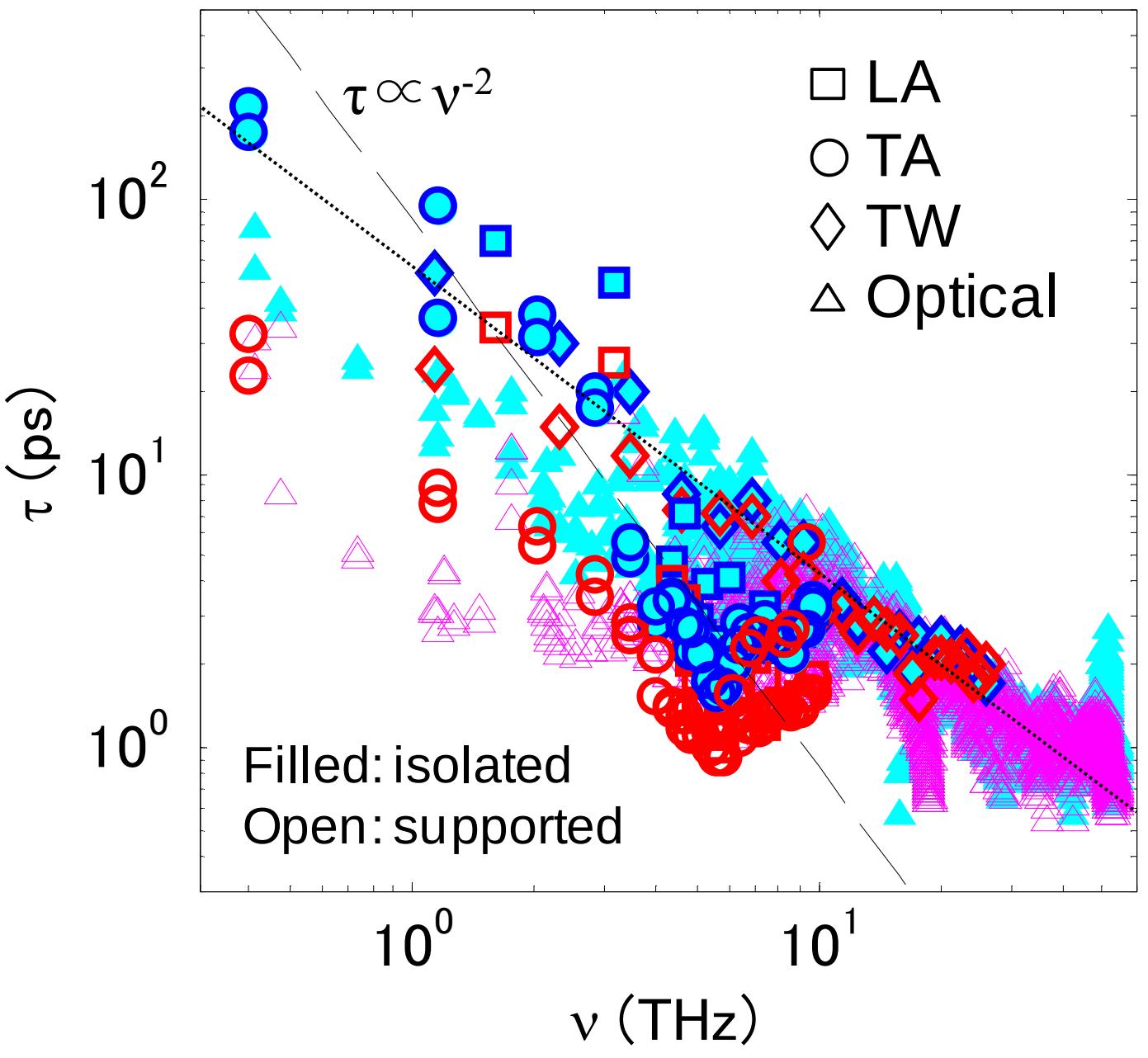

**Figure 3**: (Color online) Frequency dependence of phonon relaxation time of an isolated and supported (10,10) CNT. The data are assigned to longitudinal acoustic (LA), transverse acoustic (TA), twisting (TW), and optical phonons. The polarizations of optical phonons are not indicated for the sake of visual simplicity. The dashed line indicates $\tau \propto \nu^{-2}$ of Klemens [47]. The dotted line shows a power law ($\tau \propto \nu^{-1.1}$) fitted to the data of TW phonons in the linear dispersion regime ($\nu$ < 15 THz). The TA modes (circle symbols) show the strongest lifetime reduction between suspended and supported CNTs. Filled symbols are used for the isolated CNT, open symbols for the $SiO_2$-supported case.

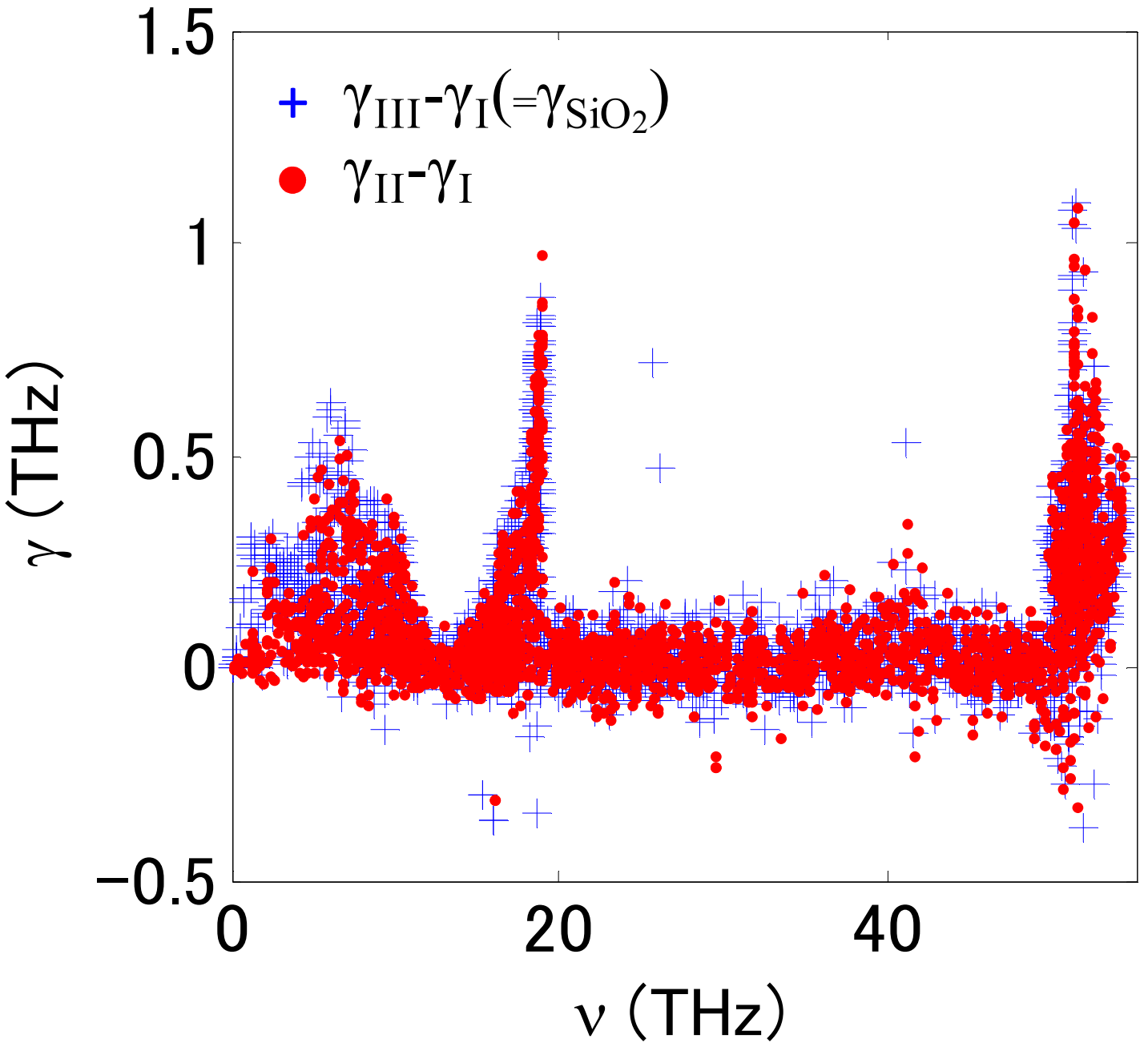

**Figure 4:** (Color online) Frequency dependence of substrate scattering rate with ($\gamma_{\mathrm{III}}$-$\gamma_{\mathrm{I}}$) and without ($\gamma_{\mathrm{II}}$-$\gamma_{\mathrm{I}}$) molecular dynamics of the a-$SiO_2$ substrate. The noise in the data and the consequent unphysical negative $\gamma$ are mainly due to the peak assignment error, which is larger for less dispersive phonons.

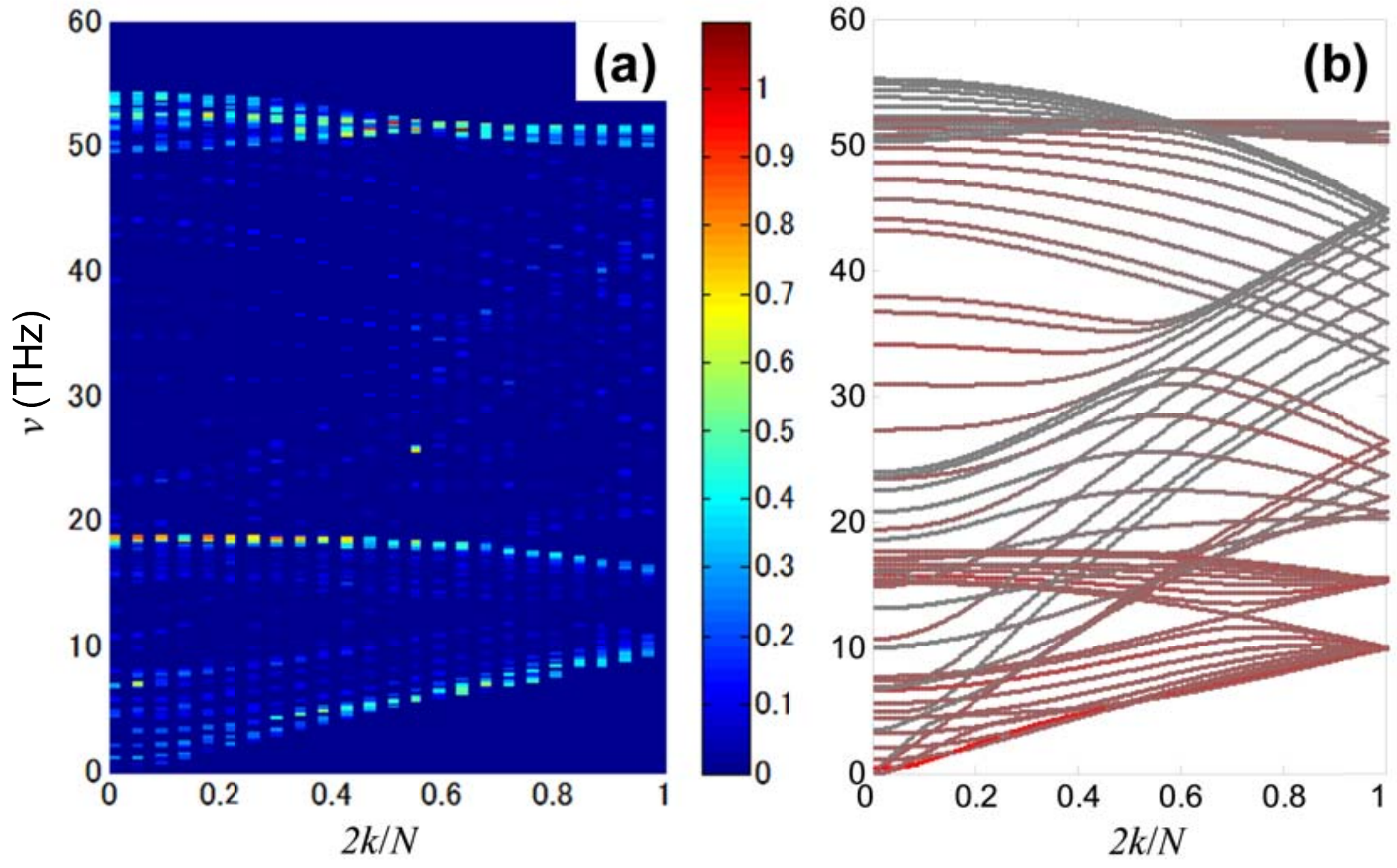

**Figure 5:** (Color online) (a) The substrate scattering rate in the ($v$,$k$)-space. The contour shows the total substrate scattering rate ($\gamma_{SiO2}$) in THz. (b) The magnitude of radial displacements of CNT eigenvectors. Each eigenvector ($k$,$k_\theta$,$\beta$) is represented by a point in ($v$,$k$) space and is colored between silver (minimum radial displacement) and red (maximum radial displacement).

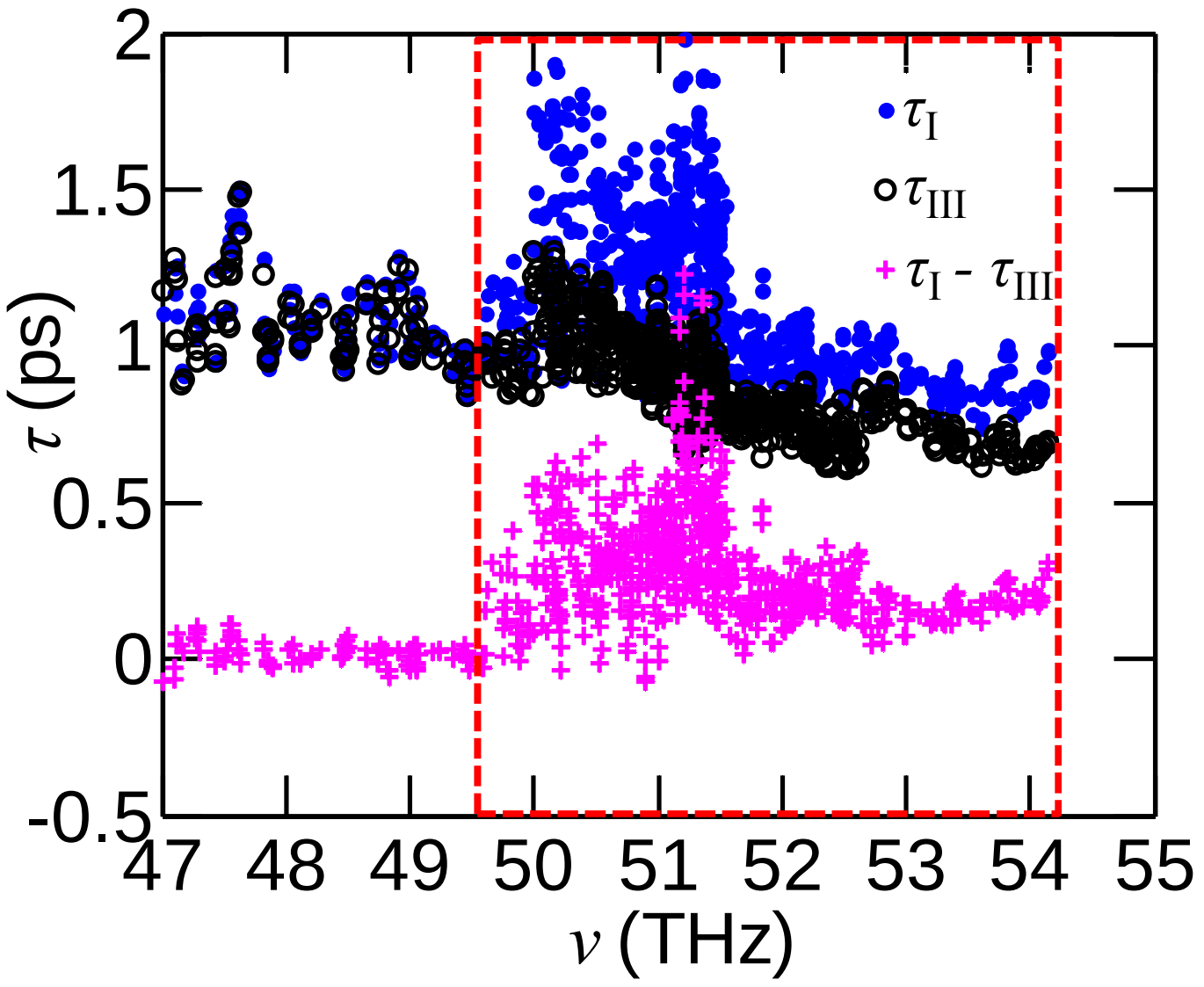

**Figure 6**: (Color online) We plot the phonon lifetimes for the isolated CNT (case I, blue dots) and the CNT on $SiO_2$ (case III, hollow black circles) in the 47 to 55 THz region, at 300 K. The window between 49.5 and 54.2 THz (enclosed by red dashed lines) contains the *G*-mode (Raman active and inactive) and other high frequency in-plane optical phonons. The lifetimes of these points are between 0.7-2 ps for the suspended CNT. Contact with the substrate leads to a significant reduction in their lifetimes, to the range $\tau_{III}$ = 0.6–1.3 ps. The difference in phonon lifetimes (plotted with magenta crosses) shows that contact with the substrate can lead to a lifetime reduction of up to 1.2 ps for phonons near the *G*-mode frequency.

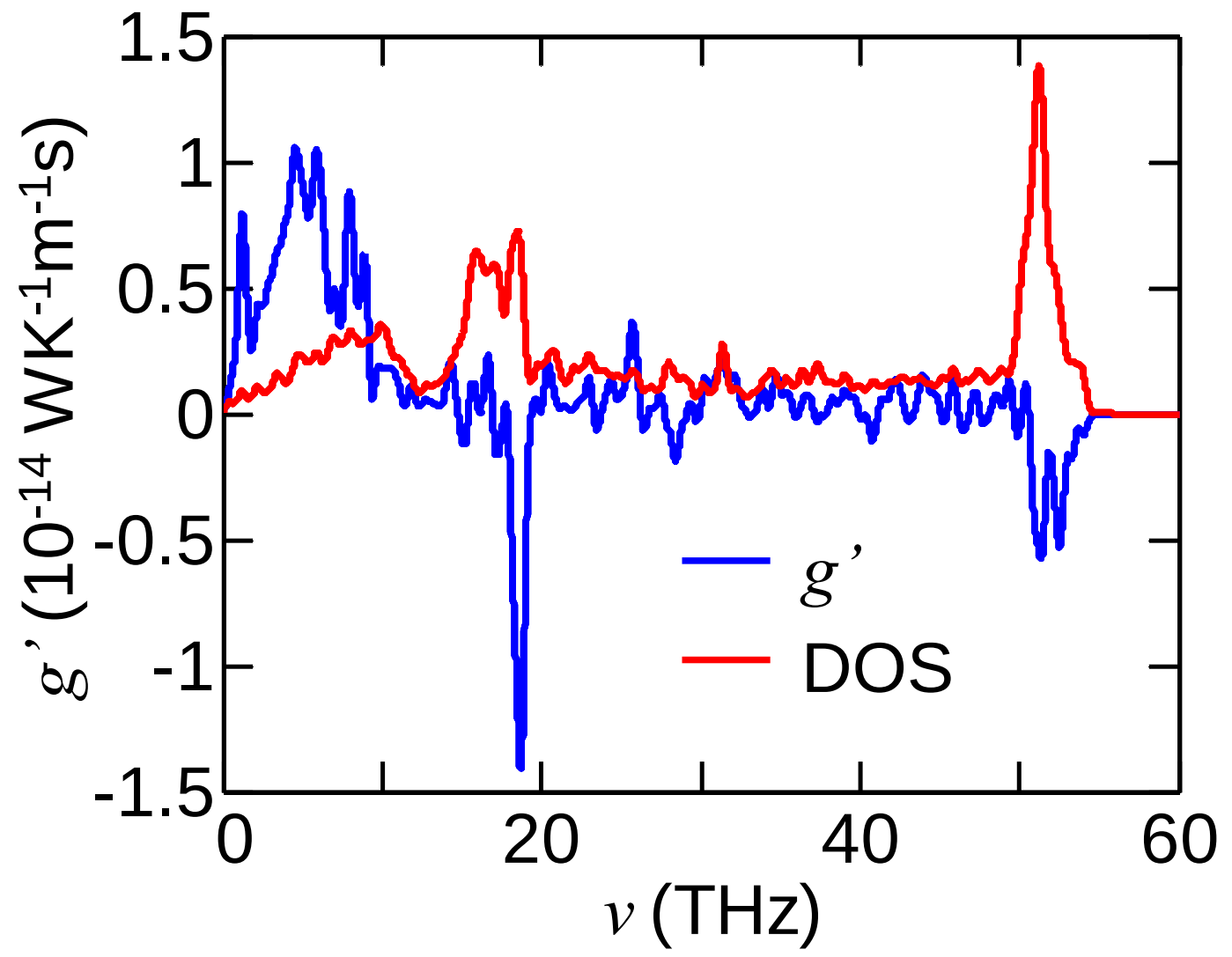

**Figure 7:** (Color online) We plot thermal boundary conductance $g'(\nu)$ and the phonon density of states (DOS) of the CNT. The negative $g'(\nu)$ values are unphysical and indicate that the CNT-substrate phonon-phonon scattering rates cannot be accurately determined from $\gamma_{\mathrm{III}}$-$\gamma_{\mathrm{II}}$. The large positive $g'(\nu)$ values between 0 and 9 THz indicate that interfacial thermal transport is dominated by phonons in this frequency range [29].

| Branch | Isolated ($Wm^{-1}K^{-1}$) | Supported ($Wm^{-1}K^{-1}$) | Reduction (%) |
|---|---|---|---|
| all | 475 | 317 | 33 |
| LA | 65 | 32 | 51 |
| TA | 71 | 12 | 83 |
| TW | 53 | 34 | 35 |
| Others | 286 | 239 | 16 |

**Table 1:** Influence of a-$SiO_2$ substrate on the overall and branch-dependent thermal conductivity of a (10,10) CNT, between the freely suspended (isolated) and substrate-supported case.